\documentclass[11pt,twoside]{article}
\usepackage[cp1251]{inputenc}  
\usepackage{graphicx}
\usepackage{amsmath}
\usepackage{amssymb}
\usepackage{mathrsfs}
\usepackage{bm} 

\begin{document}

\thispagestyle{plain}

\newcommand{\pst}{\hspace*{1.5em}}
\newcommand{\be}{\begin{equation}}
\newcommand{\ee}{\end{equation}}
\newcommand{\ds}{\displaystyle}
\newcommand{\bdm}{\begin{displaymath}}
\newcommand{\edm}{\end{displaymath}}
\newcommand{\bea}{\begin{eqnarray}}
\newcommand{\eea}{\end{eqnarray}}
\newcommand{\mbf}{\mathbf}

\begin{center} {\Large \bf
\begin{tabular}{c}
Gauge transformation of quantum states in probability representation
\end{tabular}
 } \end{center}

\smallskip

\begin{center} {\bf Ya. A. Korennoy, V. I. Man'ko}\end{center}

\smallskip

\begin{center}
{\it P.N.    Lebedev Physical Institute,                          \\
       Leninskii prospect 53, 119991, Moscow, Russia }
\end{center}

\begin{abstract}\noindent
The gauge invariance of the evolution equations of tomographic
probability distribution functions of quantum particles in an 
electromagnetic field is illustrated.
Explicit expressions for the transformations of ordinary tomograms 
of states under a gauge transformation of electromagnetic field 
potentials  are obtained. 
Gauge-independent optical and symplectic 
tomographic quasi-distributions and tomographic probability
distributions of states of quantum system are introduced, and their evolution equations 
having the Liouville equation in corresponding representations 
as the classical limit  are found.
\end{abstract}


\section{\label{Introduction}Introduction}
Gauge invariance is a fundamental quality of  classical field theory 
and quantum electrodynamics \cite{LandauII,LandauIV}, as well as of Yang-Mills theory \cite{YangMills}. 
In quantum mechanics the gauge transformation  makes the specific change \cite{LandauIII}
of the wave function \cite{Schrodinger} phase. 

For the formulation of quantum mechanics in phase space
many scientists suggested different kinds of quasidistributions
to represent the quantum states. For example, Wigner function
\cite{Wigner32},  Blohintsev function \cite{Blohintsev}, Glauber-Sudarshan $P-$function 
\cite{Glauber63,Sudarshan63}, Husimi $Q-$function \cite{Husimi40}
can be effectively used to formulate the quantum evolution and obtain
the energy levels of quantum states. All these quasidistributions are related
to wave function or density matrix by integral transformations.

In Ref.~\cite{Mancini96} the probability representation of quantum mechanics 
was introduced (see, e.g. \cite{IbortPhysScr}), in which the quantum states 
are described by fair probabilities, called quantum tomograms.
Different kinds of tomograms, e.g., optical tomograms
\cite{BerBer,VogRis}, symplectic tomograms \cite{Mancini95},
spin-tomograms \cite{OlgaJETP97,DodonovManko97} give the realization 
of star-product quantization schemes based on existence of spesific 
quantizer and dequantizer operators \cite{JPA2002,MankoMarmoVitale}.
The star-product schemes bring about the constructions for non-commutative algebra 
of  Wigner-Weil symbols of operators acting on a Hilbert space 
(see, e.g., \cite{Stratonovich1,SIGMA10086,Zachos}).

The evolution equation and energy spectrum equation for optical tomogram
were obtained in \cite{KorJRLR3274}. The evolution equation for symplectic 
tomogram was obtained in \cite{Mancini96}. On the other hand, the gauge
properties known for Schr\"odinger equation for wave function and Moyal 
equation for the Wigner function \cite{Stratonovich2} have not been considered till
now in the tomographic representation of quantum mechanics.

The aim of our paper is to explore the gauge properties of quantum tomograms,
including the star-product aspects, to introduce the gauge-independent tomograms,
and to obtain evolution equations of quantum states in gauge-independent tomographic
representations.

To begin with let us recall how the gauge invariance of non-relativistic
quantum mechanics is realized  in the wave function or density matrix
representation, and remind basic formulas of conversion of wave function and density matrix
of a particle under the gauge transformation of the potentials of the electro-magnetic field.

Consider the motion of a quantum particle having a spin in the 
electromagnetic field with the vector potential $\mathbf{A}(\mathbf{q},t)$
and the scalar potential $\varphi(\mathbf{q},t)$. 
As it is known, the Hamiltonian of such a system has the form \cite{LandauIII}
\be			\label{Hamiltonian}
\hat H=\frac{1}{2m}\left(\hat{\mathbf P}-\frac{e}{c}\mathbf{A}\right)^2
+e\varphi- \hat{\bm\varkappa}  \mathbf{B},
\ee
where $\hat{\mathbf P}=-i\hbar\partial/\partial\mathbf{q}$ is a generalized momentum operator,
$m$ and $e$ are mass and charge of the particle, $\mathbf{B}=\mathrm{rot}\mathbf{A}$ 
is a magnetic field strength, $\hat{\bm\varkappa}$ is an operator 
of quantum-mechanical  magnetic moment
\be			\label{Moment}
\hat{\bm\varkappa}=\frac{\varkappa}{s}\hat{\mathbf{s}},
\ee
where $s$ is a spin of the particle, $\hat{\mathbf{s}}$ is a spin operator,
and $\varkappa$ is a constant characteristic of the particle 
(the value of the intrinsic magnetic moment) that is the highest 
possible modulo value  $\varkappa_z$ of projection of the magnetic moment 
on the $z$ axis achieved with the projection of the spin 
on this axis equal to $s$.

From the classical electrodynamics it is known that 
potentials of the field are defined only up to the gauge transformation
\cite{LandauII}
\be			\label{eq3}
\mathbf{A} ~~\rightarrow ~~\mathbf{A} +\nabla \chi,
~~~~
\varphi~~\rightarrow ~~\varphi-\frac{1}{c}\frac{\partial\chi}{\partial t},
\ee
where $\chi$ is an arbitrary function of spatial coordinates and time.

Since the electric field intensity $\mathbf{E}$ and the magnetic field strength
$\mathbf{B}$ are defined in terms of the potentials as:
\be			\label{eq4}
\mathbf{E}=-\mathrm{grad}\varphi-\frac{1}{c}\frac{\partial}{\partial t}\mathbf{A},
~~~~
\mathbf{B}=\mathrm{rot}\mathbf{A},
\ee
then the gauge transformation (\ref{eq3}) does not affect the values 
of $\mbf E$ and $\mbf B$.
Therefore the part of Hamiltonian (\ref{Hamiltonian}) responsible for
the interaction of the spin with the magnetic field is independent 
on the gauge transformation.

The requirement of invariance of the Schr\"odinger equation 
under the gauge transformation simultaneously with the gauge-independence
of ``probability density''\, $|\Psi|^2$ leads us to the form of the conversion
of the wave function \cite{LandauIII}:
\be			\label{eq5}
\Psi ~~\rightarrow ~~\Psi \exp\left(\frac{ie}{c\hbar}\chi \right).
\ee
Accordingly, the conversions of the density matrix of the state and the Hamiltonian of the 
system under the gauge transformation acquire the forms:
\be			\label{eq6}
\hat\rho_{\mathrm c}=
\exp\left(\frac{ie}{c\hbar}\chi \right)\hat\rho\,\exp\left(-\frac{ie}{c\hbar}\chi \right),
\ee
\be			\label{eq6withstar}
\hat H_{\mathrm c}=
\exp\left(\frac{ie}{c\hbar}\chi \right)\hat H\,\exp\left(-\frac{ie}{c\hbar}\chi \right),
\ee
and the von-Neumann equation is also invariant under transformations (\ref{eq3}), (\ref{eq6})
\be			\label{eq7}
i\hbar\frac{\partial}{\partial t}\hat\rho=\left[\hat H,\hat\rho\right]
~~\rightarrow ~~
i\hbar\frac{\partial}{\partial t}\hat\rho_{\mathrm{c}}=\left[\hat H_{\mathrm c},\hat\rho_{\mathrm c}\right].
\ee

The paper is organized as follows. 
In Sec.~\ref{Sec2} we find transformations of ordinary quantum tomograms 
in general case in terms of gauge-independent quantizer and dequantizer operators.
In Sec.~\ref{Sec3} we obtain the evolution equations for classical
and quantum particles in the classical electro-magnetic field
in tomographic representations with gauge-independent dequantizers,
we discuss the gauge invariance of these equations and illustrate 
that the quantum tomographic equations 
do not be transformed to the corresponding classical equations when 
$\hbar\to0$.
In Sec.~\ref{Sec4} we introduce gauge-independent  optical and symplectic  quantum
tomographic quasidistributions and derive evolution equations for such representations.
In Sec.~\ref{Sec5} we introduce and study the gauge-independent tomographic 
probability representation and find the evolution equation for it.
Conclusion is presented in Sec.~\ref{Sec6}.

\section{\label{Sec2}Gauge transformations of ordinary quantum tomograms}
In the probability representation of quantum mechanics the states of the system
are described by a probability distribution functions $w(z,\eta,t)$ 
called quantum tomograms,
where $z$ is a set of distribution variables, $\eta$ is a set of 
parameters of corresponding tomography, and $t$ is time.
According to the universal star-product scheme (see \cite{SIGMA10086}),
the tomograms are introduced as the average values of dequantizer operators
$\hat U(z,\eta)$, 
\be			\label{eq8}
w(z,\eta,t)= \mathrm{Tr}\left\{\hat\rho(t)\hat U(z,\eta)\right\},
\ee
The inverse transformation is determined by the quantizer operator
$\hat D(z,\eta)$
\be			\label{eq9}
\hat\rho(t)=\int\hat D(z,\eta)w(z,\eta,t) dzd\eta.
\ee
The von-Neumann equation in the tomographic representation has the form
\cite{Korarticle1}:
\be			\label{eq10}
\frac{\partial }{\partial t}w(z,\eta,t)=\frac{2}{\hbar}\int\mathrm{Im}
\left[\mathrm{Tr}\left\{\hat H(t)\hat D(z',\eta'\,)\hat U(z,\eta)\right\}\right]
w(z',\eta'\,,t)dz'd\eta'.
\ee
It is easy to see that if  we determine in definition (\ref{eq8}) that the dequantizer 
and the quantizer are gauge-independent, 
then equation (\ref{eq10}) is invariant under the gauge transformation 
only with the following  transformation of tomograms:
\bea			
w(z,\eta,t)&\rightarrow&
w_\mathrm{c}(z,\eta,t)=\mathrm{Tr}\left\{
\exp\left(\frac{ie}{c\hbar}\chi\right)\hat\rho(t)\exp\left(-\frac{ie}{c\hbar}\chi \right)
\hat U(z,\eta)
\right\} \nonumber \\
&&=\mathrm{Tr}\left\{
\exp\left(\frac{ie}{c\hbar}\chi \right)\int\hat D(z',\eta'\,)w(z',\eta',t)dz'd\eta'
\exp\left(-\frac{ie}{c\hbar}\chi \right)
\hat U(z,\eta)
\right\}.
\label{eq11}
\eea
Introducing the notation for the kernel $G(z,\eta,z',\eta'\,)$
\be			\label{eq11_1}
G(z,\eta,z',\eta'\,)=\mathrm{Tr}\left\{
\exp\left(\frac{ie}{c\hbar}\chi \right)\hat D(z',\eta'\,)
\exp\left(-\frac{ie}{c\hbar}\chi \right)
\hat U(z,\eta)
\right\},
\ee
for the gauge transformed function $w_{\mathrm c}(z,\eta,t)$ we get
\be			\label{eq11_2}
w_\mathrm{c}(z,\eta,t)=\int G(z,\eta,z',\eta'\,)w(z',\eta',t)dz'd\eta'.
\ee
Thus, under the gauge transformation of the electromagnetic field potentials 
the tomogram of the state is converted by means of integral transformation 
(\ref{eq11_2}), in which the explicit form of the kernel depends on the 
type of tomography.

If we have spinless quantum particle with mass $m$ in three-dimensional space,
then the dequantizer and the quantizer for the optical tomography
have the form \cite{KorPhysRevA85}
\be		\label{dequantizerOPT}
\hat U_w(\mbf X,\bm\theta)=\prod_{\sigma=1}^3
\delta\left(X_\sigma-\hat q_\sigma\cos\theta_\sigma-\hat P_\sigma\frac{\sin\theta_\sigma}
{m\omega_{\sigma}}\right),
\ee
\be		\label{quantizerOPT}
\hat D_w(\mbf X,\bm\theta)=\int\prod_{\sigma=1}^3
\frac{\hbar\vert\eta_\sigma\vert}{2\pi m\omega_{\sigma}}
\exp\left\{i\eta_\sigma\left(X_\sigma-\hat q_\sigma\cos\theta_\sigma
-\hat P_\sigma\frac{\sin\theta_\sigma}
{m\omega_{\sigma}}\right)\right\}
d^3\eta,
\ee
where $\hat P_\sigma$ are components of the generalized momentum operator and
$\omega_\sigma$ are constants that have the dimension
of frequency. Further for simplicity we choose the set $\{\omega_\sigma\}$ so that
$\omega_1=\omega_2=\omega_3=\omega$.

Substituting these expressions of the dequantizer and the quantizer 
to equation (\ref{eq11_1}), after some calculations 
using the formula for the matrix elements
\be			\label{eq13}
\langle q_\sigma'|e^{i(a\hat q_\sigma+b\hat P_\sigma)}|q_\sigma\rangle=
e^{ia(q_\sigma+q_\sigma')/2}\delta(q_\sigma'-q_\sigma+b),
\ee
we obtain
\bea
G_w(\mbf{X},\bm{\theta},\mbf{X}',\bm{\theta}\,')&=&\frac{1}{(4\pi^2\hbar)^3}\int
\exp\left\{\frac{ie}{c\hbar}
\left[\chi\left(\frac{k_\sigma\sin\theta_\sigma}{m\omega}+
\frac{\sqrt{\hbar}\sin\theta_\sigma'}{2\sqrt{m\omega}}r_\sigma\right)
-\chi\left(\frac{k_\sigma\sin\theta_\sigma}{m\omega}-
\frac{\sqrt{\hbar}\sin\theta_\sigma'}{2\sqrt{m\omega}}r_\sigma\right)\right]
\right\} \nonumber \\
&\times&
\prod_{\sigma=1}^3|r_\sigma|\exp\left\{
ir_\sigma\sqrt{\frac{m\omega}{\hbar}}\left(X_\sigma'-
X_\sigma\frac{\sin\theta_\sigma'}{\sin\theta_\sigma}\right)
-ik_\sigma r_\sigma\frac{\sin(\theta-\theta'\,)}{\sqrt{m\omega\hbar}}
\right\}d^3k\,d^3r.
\label{eq16}
\eea
Further simplification of this expression can, unfortunately, possible only with
the explicit expression for the function $\chi$.

For the spinless symplectic tomography dequantizer and quantizer are given by formulas
\be		\label{dequantizerSYMP}
\hat U_M(\mbf X,\bm\mu,\bm\nu)=\prod_{\sigma=1}^3
\delta(X_\sigma-\hat q_\sigma\mu_\sigma-\hat P_\sigma\nu_\sigma),
\ee
\be		\label{quantizerSYMP}
\hat D_M(\mbf X,\bm\mu,\bm\nu)=
\prod_{\sigma=1}^3\frac{m\omega}{2\pi}
\exp\left\{i\sqrt{\frac{m\omega}{\hbar}}
\left(X_\sigma-\hat q_\sigma\mu_\sigma-\hat P_\sigma\nu_\sigma\right)\right\},
\ee
and from (\ref{eq11_1}) we can obtain
\bea			
G_M(\mbf{X},\bm{\mu},\bm{\nu},\mbf{X}',\bm{\mu}',\bm{\nu}'\,)&=&
\left(\frac{m\omega}{4\pi^2\hbar}\right)^3
\int\exp\left\{
\frac{ie}{c\hbar}\left[\chi\left(\nu_\sigma k_\sigma + 
\frac{\sqrt{m\omega\hbar}}{2}\nu\,'\!\!\!_\sigma\right)
-\chi\left(\nu_\sigma k_\sigma-\frac{\sqrt{m\omega\hbar}}{2}\nu\,'\!\!\!_\sigma\right)\right]
\right\}
\nonumber \\
&&\times\prod_{\sigma=1}^3
\exp\left\{i\sqrt{\frac{m\omega}{\hbar}}\left[k_\sigma(\mu\nu\,'\!\!\!_\sigma-\mu_\sigma'\nu)
+X_\sigma'-X_\sigma\frac{\nu\,'\!\!\!_\sigma}{\nu_\sigma}\right]\right\}d^3k.
\label{eq14}
\eea
Note, that the kernels $G_w$ and $G_M$ are connected by the relation
\bdm
G_w(\mbf{X},\bm{\theta},\mbf{X}',\bm{\theta}\,')=\int\frac{|r_1|\,|r_2|\,|r_3|}{(m\omega)^3}
G_M\left(X_\sigma,\cos\theta_\sigma,\frac{\sin\theta_\sigma}{m\omega},
r_\sigma X',r_\sigma\cos\theta_\sigma',r_\sigma\frac{\sin\theta_\sigma'}{m\omega}\right)
d^3r.
\edm

Consider now the positive vector non-redundant tomography of the particle with spin
\cite{Korarticle9}, \cite{Korarticle14}
\be			\label{deffn21}
\mbf w(z,\eta,t)=\mbox{Tr}\left\{
\hat\rho(t)\hat{\mbf U}(z,\eta)
\right\},
\ee
where the trace is calculated also over spin indexes.
In this representation the components of the dequantizer and the quantizer are
defined by formulas
\be			\label{eq17}
\hat U_{j(nl)}(z,\eta)=\hat U(z,\eta)\otimes\hat{\mathcal{U}}_{j(nl)},
~~~~
\hat D_{(nl)j}(z,\eta)=\hat D(z,\eta)\otimes\hat{\mathcal{D}}_{(nl)j},
\ee
where $\hat{\mathcal{U}}_{j(nl)}$ and $\hat{\mathcal{D}}_{(nl)j}$ are 
spin dequantizer and quantizer, $j=\overline{1,\,(2s+1)^2}$ is the index corresponding 
to the $j$th component of the vector tomogram $\mbf w(z,\eta,t)$, and
$n,l=\overline{1,\,(2s+1)}$ are spin indexes.
Since
\be			\label{eq18}
\sum_{n,l=1}^{2s+1}\hat{\mathcal{D}}_{(nl)j'}\,\hat{\mathcal{U}}_{j(ln)}
=\delta_{jj'},
\ee
then, according to general formula (\ref{eq11_1}), the kernel of the
transformation of the vector tomogram $\mbf w(z,\eta,t)$ takes the form:
\be			\label{eq19}
G_{jj'}(z,\eta,z',\eta')=
\mathrm{Tr}\left\{
\exp\left(\frac{ie}{c\hbar}\chi \right)\hat D(z',\eta'\,)\otimes\hat{\mathcal{D}}_{j'}
\exp\left(-\frac{ie}{c\hbar}\chi \right)
\hat U(z,\eta)\otimes\hat{\mathcal{U}}_j
\right\}
=\delta_{jj'}G(z,\eta,z',\eta'),
\ee
i.e., the vector tomogram $\mbf w(z,\eta,t)$ under the gauge transformation
is converted by components through the integral transformation:
\be			\label{eq20}
\mbf w(z,\eta,t) ~~\rightarrow ~~\mbf w_\mathrm{c}(z,\eta,t) =
\int G(z,\eta,z',\eta'\,)\mbf w(z',\eta',t)dz'd\eta',
\ee
where $G(z,\eta,z',\eta')$ is a kernel  of the integral transformation
for the spinless case.
This formula is valid for arbitrary spin.

Thus we see that if the dequantizer is gauge-independent, then the tomogram is gauge-dependent,
and the evolution equation is gauge-invariant but gauge-dependent.

\section{\label{Sec3}Gauge invariance of evolution equations}
Let us consider in more detail the gauge invariance of the quantum 
evolution equations in the tomographic representations with 
gauge-independent dequantizers and the question of limiting transition 
of such equations to classics when $\hbar\to0$.
At first, we will get the Liouville equation in the 
electro-magnetic field in the tomographic representations.

For the classical ensemble of non-interacting particles with 
mass $m$ and charge $e$ this equation in the phase space has the form:
\be			\label{Liouville}
\frac{\partial}{\partial t}W_\mathrm{cl}(\mathbf{q},\mathbf{p},t)+
\frac{\mathbf{p}}{m}\frac{\partial}{\partial\mathbf{q}}W_\mathrm{cl}(\mathbf{q},\mathbf{p},t)
+e\left(\mathbf{E}(\mathbf{q},t)
+\frac{1}{mc}[\mathbf{p}\times\mathbf{B}(\mathbf{q},t)]\right)
\frac{\partial}{\partial\mathbf{p}}W_\mathrm{cl}(\mathbf{q},\mathbf{p},t)=0,
\ee
where $\mathbf{p}$ is a kinetic momentum, $\mathbf{E}(\mathbf{q},t)$ and 
$\mathbf{B}(\mathbf{q},t)$ are electric and magnetic fields, 
defined by the formulas (\ref{eq4}), $W_\mathrm{cl}(\mathbf{q},\mathbf{p},t)$
is a distribution function of non-interacting particles.

The distribution function $W_\mathrm{cl}(\mathbf{q},\mathbf{p},t)$ is independent on 
the geauge transformation \cite{LandauII}, 
because the Liouville equation (\ref{Liouville}) includes 
only gauge-independent intensities of the electro-magnetic field.
Consequently, the optical and symplectic tomograms of the function
$f(\mathbf{q},\mathbf{p},t)$ defined by the formulas \cite{KorJRLR3274} 
\be		\label{eq_50}
w_\mathrm{cl}(\mbf x,\bm\theta,t)=\int W_\mathrm{cl}(\mathbf{q},\mathbf{p},t)
\prod_{\sigma=1}^{3} \delta\left(x_\sigma
-q_\sigma\cos\theta_\sigma-p_\sigma\frac{\sin\theta_\sigma}{m\omega}\right)
d^3q\,d^3p,
\ee
\be		\label{eq_51}
M_\mathrm{cl}(\mbf x,\bm\mu,\bm\nu,t)=\int W_\mathrm{cl}(\mathbf{q},\mathbf{p},t)
\prod_{\sigma=1}^{3}\delta\left(x_\sigma-\mu_\sigma q_\sigma
-\nu_\sigma p_\sigma\right)d^3q\,d^3p,
\ee
are also independent on the gauge transformation. We use the designation $\mbf x$
instead of $\mbf X$ for distribution variable to point out that the Radon
transformations (\ref{eq_50}) and (\ref{eq_51}) are made in the phase space 
with kinetic momentum $\mbf p$.

Using the known correspondence rules \cite{KorJRLR3274,KorPhysRevA85}
between the operators acting on the Wigner function \cite{Wigner32} 
(or on the distribution function) and the operators acting on the 
optical or symplectic tomograms
\be		\label{CorrespRules}
\begin{array} {lclcl} 
q_\sigma W(\mathbf{q},\mathbf{p}) &\longleftrightarrow &
-\partial_{x_\sigma}^{-1}\partial_{\mu_\sigma}M(\mbf x,\bm\mu,\bm\nu)
&\longleftrightarrow &
\left(\sin\theta_\sigma\partial_{x_\sigma}^{-1}\partial_{\theta_\sigma}
+x_\sigma\cos\theta_\sigma\right)w(\mbf x,\bm\theta),
\\
p_\sigma W(\mathbf{q},\mathbf{p}) &\longleftrightarrow &
-\partial_{x_\sigma}^{-1}\partial_{\nu_\sigma}M(\mbf x,\bm\mu,\bm\nu)
&\longleftrightarrow &
m\omega\left(-\cos\theta_\sigma\partial_{x_\sigma}^{-1}\partial_{\theta_\sigma}
+x_\sigma\sin\theta_\sigma\right)w(\mbf x,\bm\theta),
\\
\partial_{q_\sigma} W(\mathbf{q},\mathbf{p}) &\longleftrightarrow &
\mu_\sigma\partial_{x_\sigma}M(\mbf x,\bm\mu,\bm\nu)
&\longleftrightarrow &
\cos\theta_\sigma\partial_{x_\sigma}w(\mbf x,\bm\theta),
\\
\partial_{p_\sigma} W(\mathbf{q},\mathbf{p}) &\longleftrightarrow &
\nu_\sigma\partial_{x_\sigma}M(\mbf x,\bm\mu,\bm\nu)
&\longleftrightarrow &
\ds\frac{\sin\theta_\sigma}{m\omega}\partial_{x_\sigma}w(\mbf x,\bm\theta),
\end{array}
\ee
where we introduced the designation \cite{KorPhysRevA85} for inverse derivatives
\be			\label{invder}
\partial_{x_\sigma}^{-n}f(x_\sigma)=\frac{1}{(n-1)!}
\int(x_\sigma-x_\sigma')^{n-1}\Theta(x_\sigma-x_\sigma')f(x_\sigma')dx_\sigma',
\ee
where $\Theta(x_\sigma-x_\sigma')$ is a Heaviside step function,
we can  write Liouville equation (\ref{Liouville}) in the 
optical and the symplectic tomography representation
\begin{eqnarray}                             
&&\partial_t w_\mathrm{cl}(\mbf x,\bm\theta,t)=
\Bigg[\omega\sum_{j=1}^{3} \left(\cos^2\theta_j\partial_{\theta_j}
-\frac{\sin2\theta_j}{2}\Big\{1+x_j\partial_{x_j}\Big\}
\right)\nonumber \\
&&+\frac{e}{mc}\sum_{\alpha,\beta,\gamma=1}^3 \varepsilon_{\alpha\beta\gamma}
B_\gamma\Big(\sin\theta_\sigma\partial_{x_\sigma}^{-1}\partial_{\theta_\sigma}
+x_\sigma\cos\theta_\sigma,t\Big)
\left(\cos\theta_\beta\partial_{x_\beta}^{-1}\partial_{\theta_\beta}
-x_\beta\sin\theta_\beta\right)\sin\theta_\alpha\partial_{x_\alpha}
\nonumber \\
&&-\frac{e}{m\omega}\sum_{j=1}^{3}E_j\Big(\sin\theta_\sigma\partial_{x_\sigma}^{-1}\partial_{\theta_\sigma}
+x_\sigma\cos\theta_\sigma,t\Big)
\sin\theta_j\partial_{x_j}\Bigg]
w_\mathrm{cl}(\mbf x,\bm\theta,t),
		\label{LivTomOpt}
\end{eqnarray}
\begin{eqnarray}
\partial_tM_\mathrm{cl}(\mbf x,\bm\mu,\bm\nu,t)&=&
\left[\frac{\bm\mu}{m}\partial_{\bm\nu}
+\frac{e}{mc}\sum_{\alpha,\beta,\gamma=1}^3 \varepsilon_{\alpha\beta\gamma}
B_\gamma\Big(-\partial_{x_\sigma}^{-1}\partial_{\mu_\sigma},t\Big)
\left(\partial_{x_\beta}^{-1}\partial_{\nu_\beta}\right)\nu_\alpha\partial_{x_\alpha}\right.
\nonumber \\
&&-\left.e\sum_{j=1}^{3}
E_j\Big(-\partial_{x_\sigma}^{-1}\partial_{\mu_\sigma},t\Big)\nu_j\partial_{x_j}
\right]M_\mathrm{cl}(\mbf x,\bm\mu,\bm\nu,t),
		\label{LivTomSym}
\end{eqnarray}
where $\varepsilon_{\alpha\beta\gamma}$ is the completely antisymmetric 
pseudo-tensor of 3rd rank (the Levi-Civita symbol).

Thus, we have gauge-independent equations (\ref{LivTomOpt},\ref{LivTomSym})
for gauge-independent classical tomograms $w_\mathrm{cl}(\mbf x,\bm\theta,t)$ and
$M_\mathrm{cl}(\mbf x,\bm\mu,\bm\nu,t)$.

As it is known, if we have the ensemble of non-interacting particles
in the potential field, the generalized momentum of the particle is equal 
to its kinetic momentum, and the quantum analogue of the Liouville equation
in this case is Moyal equation \cite{Moyal1949} for the Wigner function \cite{Wigner32}
\be			\label{WigDef}
W(\mathbf{q},\mathbf{P},t)=\frac{1}{(2\pi\hbar)^3}
\int\rho\left(\mathbf{q}-\frac{\mathbf u}{2},\mathbf{q}+\frac{\mathbf u}{2},t\right)
\exp\left(\frac{i}{\hbar}\mathbf{u}\mathbf{P}\right)d^3u,
\ee
which is converted into the Liouville equation when $\hbar\to0$.

In the electro-magnetic field when  $\mathbf{A}\not=0$, 
the Moyal equation for the function (\ref{WigDef}) is written as follows:
\bea			
\frac{\partial }{\partial t} W({\mathbf q},{\mathbf P},t)&=&
\left[-\frac{\mathbf P}{m}
\frac{\partial}{\partial{\mathbf q}}
+\frac{2e}{\hbar}\,\mbox{Im}\,{\varphi}
\left(
{\mathbf q}+\frac{i\hbar}{2} \frac{\partial}{\partial {\mathbf P}},t
\right)
+\frac{e^2}{mc^2\hbar}\,\mathrm{Im}{\bf A}^2
\left(
{\mathbf q}+\frac{i\hbar}{2} \frac{\partial}{\partial {\mathbf P}},t
\right)\right.
 \nonumber \\
&-&\left.
\frac{2e}{mc\hbar}\,{\mathrm{Im}}\left\{{\mathbf A}
\left(
{\mathbf q}+\frac{i\hbar}{2} \frac{\partial}{\partial {\mathbf P}},t
\right)
\left(
{\mathbf P}-\frac{i\hbar}{2} \frac{\partial}{\partial {\mathbf q}}
\right)
\right\}\right.  \nonumber \\
&+&\left.\frac{e}{mc}\,\mathrm{Re}{\nabla_{\bf q}{\bf A}}
\left(
\mathbf q \to {\mathbf q}+\frac{i\hbar}{2} \frac{\partial}{\partial {\mathbf P}},t
\right)
 \right] W({\mathbf q},{\mathbf P},t).
\label{MoyalOldWig}
\eea
The function $W({\mathbf q},{\mathbf P},t)$ is gauge-dependent, 
but if we take the classical limit $\hbar\to0$ 
and change variables $\mathbf{p}=\mathbf{P}-\frac{e}{c}\mathbf{A}$ 
in  (\ref{MoyalOldWig}), then
this equation will be converted into gauge-independent Liouville equation (\ref{Liouville}).
However, there is no contradiction here, because in the gauge transformation
of the function \mbox{$W(\mbf q,\mbf p+\frac{e}{c}\mbf A)$}
\bea
W_{\mathrm c}\left(\mbf q,\mbf p+\frac{e}{c}\mbf A\right)&=&
\int W\left(\mbf q,\mbf p'+\frac{e}{c}\mbf A\right)
\exp\left\{\frac{i}{\hbar}\mbf u(\mbf p-\mbf p')\right\}
\nonumber \\
&\times&
\exp\left\{\frac{ie}{c\hbar}\left[\chi\left(\mbf q-\frac{\mbf u}{2}\right)
-\chi\left(\mbf q+\frac{\mbf u}{2}\right)+\mbf u\nabla\chi(\mbf q) \right]\right\}
\frac{d^3u\,d^3p'}{(2\pi\hbar)^3}
\nonumber
\eea
we can spread out the function $\chi(\mbf q\pm\mbf u/2)$ up to the first order
$\chi(\mbf q\pm\mbf u/2)\approx\chi(\mbf q)\pm\frac{1}{2}\mbf u\nabla\chi(\mbf q)$
using a method of a stationary phase at $\hbar\to0$.
After that in the limit case we obtain
\bdm
W_{\mathrm c}\left(\mbf q,\mbf p+\frac{e}{c}\mbf A\right)=
\int W\left(\mbf q,\mbf p'+\frac{e}{c}\mbf A\right)
\delta(\mbf p-\mbf p')
d^3p'=W\left(\mbf q,\mbf p+\frac{e}{c}\mbf A\right),
\edm
that is the function $W\left(\mbf q,\mbf p+\frac{e}{c}\mbf A\right)$ becomes gauge-independent.

Let us transform Moyal equation (\ref{MoyalOldWig}) to the optical and
symplectic tomographic representations, in which the tomograms
$w(\mbf X,\bm\theta,t)$ and $M(\mbf X,\bm\mu,\bm\nu,t)$ are defined from 
the Wigner function $W({\mathbf q},{\mathbf P},t)$ with the same
formulas (\ref{eq_50}) and (\ref{eq_51}), where the kinetic momentum $\mathbf p$
should be replaced by the generalized momentum $\mathbf P$, 
and the variable $\mbf x$ should be replaced by $\mbf X$ 
to point out that the Radon transformations are being done in the 
phase space with generalized momentum.
For this purpose we should use the same correspondence rules as (\ref{CorrespRules}).
After calculations we can write the evolution equation for gauge-dependent optical tomogram
as follows:
\bea
\partial_t w(\mbf X,\bm\theta,t)&=&
\Bigg[\omega\sum_{j=1}^{3}\left(\cos^2\theta_j\partial_{\theta_j}
-\frac{1}{2}\sin2\theta_j\Big\{1+X_j\partial_{X_j}\Big\}\right) 
+\frac{2e}{\hbar}\,\mathrm{Im}\,[\hat{\varphi}]_w
\nonumber \\
&+&\frac{e^2}{mc^2\hbar}\,\mathrm{Im}[\hat{\bf A}]_w^2
-\frac{2e}{mc\hbar}\,\mathrm{Im}\left([\hat{\mathbf A}]_w [\hat{\mathbf P}]_w\right)
+\frac{e}{mc}\,\mathrm{Re}\left[{\nabla_{\mathbf q}{\mathbf A}}\right]_w\Bigg]
w(\mbf X,\bm\theta,t),
\label{EqTomOpt}
\eea
where
\bdm
[\hat A_j]_w=A_j\left([\hat{\mathbf q}]_w,t\right),
~~~~
[\hat{\varphi}]_w=\varphi\left([\hat{\mathbf q}]_w,t\right),
\edm
\bdm
\left[\nabla_{\mathbf q}\hat{\mathbf A}\right]_w
=\nabla_{\mathbf q}{\mathbf A}\left(
{\mathbf q}\rightarrow [\hat{\bf q}]_w,t\right),
\edm
and $[\hat{\mathbf q}]_w$, $[\hat{\mathbf P}]_w$ are position
and generalized momentum operators in the optical tomographic representation \cite{KorPhysRevA85},
\be			\label{qFromXTheta}
[\hat q_\sigma]_w=\sin\theta_\sigma\partial_{\theta_\sigma}
\partial_{X_\sigma}^{-1}
+X_\sigma\cos\theta_\sigma+i\frac{\hbar\sin\theta_\sigma}
{2m\omega}
\partial_{X_\sigma},
\ee
\be			\label{pFromXTheta}
[\hat P_\sigma]_w=m\omega\left(-\cos\theta_\sigma\partial_{X_\sigma}^{-1}
\partial_{\theta_\sigma}+X_\sigma\sin\theta_\sigma\right)-\frac{i\hbar}{2}
\cos\theta_\sigma\partial_{X_\sigma}.
\ee
For the gauge-dependent symplectic tomogram we can write
\bea
\partial_t M({\mbf X},{\bm\mu},\bm\nu,t)&=&\left[
\frac{\bm\mu}{m}\partial_{\bm\nu}
+\frac{2e}{\hbar}\,\mathrm{Im}\,[\hat{\varphi}]_M
+\frac{e^2}{mc^2\hbar}\,\mathrm{Im}[\hat{\mathbf A}]_M^2
\right.
\nonumber \\
&-&\left.\frac{2e}{mc\hbar}\,\mathrm{Im}\left([\hat{\mathbf A}]_M [\hat{\mathbf P}]_M\right)
+\frac{e}{mc}\,\mathrm{Re}\left[\nabla_{\mathbf q}{\mathbf A}\right]_M\right]
M({\mbf X},{\bm\mu},\bm\nu,t),
\label{EqTomSym}
\eea
where
\bdm
[\hat A_j]_M=A_j\left([\hat{\mathbf q}]_M,t\right),~~~~
[\hat\varphi]_M=\varphi\left([\hat{\mathbf q}]_M,t\right),
\edm
\bdm
[\nabla_{\mathbf q}{\mathbf A}]_M=\nabla_{\mathbf q}{\mathbf A}\left(
{\mathbf q}\rightarrow [\hat{\mathbf q}]_M,t\right), 
\edm
and $[\hat{\mathbf q}]_M$, $[\hat{\mathbf P}]_M$ are position
and generalized momentum operators in the symplectic representation (see \cite{KorJRLR3274}),
\be			\label{defqpSymp}
[\hat P_\sigma]_M=-\partial_{X_\sigma}^{-1}
\partial_{\nu_\sigma}-i(\hbar/2)\mu_\sigma\partial_{X_\sigma},
~~~~
[\hat q_\sigma]_M=-\partial_{X_\sigma}^{-1}
\partial_{\mu_\sigma}+i(\hbar/2)\nu_\sigma\partial_{X_\sigma}.
\ee
Equations (\ref{EqTomOpt}), (\ref{EqTomSym}) are gauge-invariant only under the 
condition of transformation of tomograms with general formula (\ref{eq11_2})
with the kernel $G(z,\eta,z',\eta'\,)$ defined by formula (\ref{eq16}) or (\ref{eq14}).
In the classical limit $\hbar\to0$ these equations, in general case, are not 
converted into equations (\ref{LivTomOpt}), (\ref{LivTomSym}).
The thing is that (\ref{LivTomOpt}) and (\ref{EqTomOpt})
are equations for distribution functions of different observables:
$x_\sigma(\theta_\sigma)=q_\sigma\cos\theta_\sigma+p_\sigma\sin\theta_\sigma$
in the classical case (\ref{LivTomOpt}); 
but $\hat X_\sigma(\theta_\sigma)=
\hat q_\sigma\cos\theta_\sigma+\hat P_\sigma\sin\theta_\sigma$
in the quantum case (\ref{EqTomOpt}).
Analogously,  (\ref{LivTomSym}) and (\ref{EqTomSym}) are equations 
for distribution functions of different observables
$x_\sigma(\mu_\sigma,\nu_\sigma)=q_\sigma\mu_\sigma+p_\sigma\nu_\sigma$
and $\hat X_\sigma(\mu_\sigma,\nu_\sigma)=\hat q_\sigma\mu_\sigma+
\hat P_\sigma\nu_\sigma$ respectively.

\section{\label{Sec4}Gauge-independent tomographic quasiprobability representations}
In the previous section we have shown that the  evolution equations
in the tomographic representations for the gauge-dependent tomograms in the classical limit
$\hbar\to0$ are not converted to the Liouville equation in the tomographic forms 
for gauge-independent tomograms of the classical distribution function.

Therefore, for the construction of quantum tomographic representations, 
in which the evolution equations would have been transformed
to (\ref{LivTomOpt}) and (\ref{LivTomSym}) when $\hbar\to0$, 
we need to introduce gauge-independent quantum tomograms.
This can be done with the help of a gauge-independent Wigner function
obtained in \cite{Stratonovich2},
\be			\label{WigNew}
W_\mathrm{g}(\mathbf{q},\mathbf{p},t)=\frac{1}{(2\pi\hbar)^{3}}\int
\exp\left(\frac{i}{\hbar}\mathbf{u}\left\{\mathbf{p}
+\frac{e}{c}\int_{-1/2}^{1/2} d\tau \mathbf{A}(\mathbf{q}+\tau\mathbf{u},\,t)\right\}\right)
\rho\left(\mathbf{q}-\frac{\mathbf u}{2},\,\mathbf{q}+\frac{\mathbf u}{2},\,t\right)
d^3u,
\ee
where $\mathbf{p}$ is a kinetic  momentum.

Gauge-independent Moyal equation for this function has the form \cite{Serimaa1986}:
\be			\label{EqWigNew}
\left\{\partial_t+\frac{1}{m}\left(\mathbf{p}+\triangle\tilde{\mathbf p}\right)\partial_{\mathbf{q}}
+ e\left(\tilde{\mathbf{E}} +\frac{1}{mc}
\left[\left(\mathbf{p}+\triangle\tilde{\mathbf p}\right)\times\tilde{\mathbf{B}} \right]\right)
\partial_{\mathbf p}
\right\}W_\mathrm{g}(\mathbf{q},\mathbf{p},t)=0,
\ee
where
\bdm
\triangle\tilde{\mathbf p}=-\frac{e}{c}\frac{\hbar}{i}
\left[
\frac{\partial}{\partial\mathbf{p}} \times \int_{-1/2}^{1/2}
d\tau\,\tau 
\mathbf{B}
\left(
\mathbf{q}+i\hbar\tau\frac{\partial}{\partial\mathbf{p}},\,t
\right)
\right],
\edm
\bdm
\tilde{\mathbf E}=\int_{-1/2}^{1/2}
d\tau\,
\mathbf{E}
\left(
\mathbf{q}+i\hbar\tau\frac{\partial}{\partial\mathbf{p}},\,t
\right),
~~~~
\tilde{\mathbf B}=\int_{-1/2}^{1/2}
d\tau\,
\mathbf{B}
\left(
\mathbf{q}+i\hbar\tau\frac{\partial}{\partial\mathbf{p}},\,t
\right).
\edm
This equation in the classical limit $\hbar\to 0$ is converted into 
Liouville equation (\ref{Liouville}).

If we apply Radon transformations (\ref{eq_50}) and (\ref{eq_51}) to 
the Wigner function (\ref{WigNew}), we obtain 
gauge-independent optical $w_\mathrm{g}(\mbf x,\bm\theta,t)$ and symplectic
$M_\mathrm{g}(\mbf x,\bm\mu,\bm\nu,t)$ tomograms.
Under such definitions the correspondence rules between operators 
acting on the tomograms and the Wigner function will be similar to the
correspondence rules (\ref{CorrespRules}).
Then, from equation (\ref{EqWigNew}) we find the evolution equation 
for the gauge-independent optical tomogram $w_\mathrm{g}(\mbf x,\bm\theta,t)$:
\bea
&&\partial_tw_\mathrm{g}(\mbf x,\bm\theta,t)=\Bigg[\omega\sum_{j=1}^{3}
\bigg
(\cos^2\theta_j\partial_{\theta_j}
-\frac{1}{2}\sin2\theta_j\Big\{1+x_j\partial_{x_j}\Big\}
\nonumber \\
&&-\frac{1}{m}\sum_{\alpha=1}^3\left[\triangle \tilde{\mathbf p}_\alpha\right]_{w}
\cos\theta_\alpha\partial_{x_\alpha}
-\frac{e}{m\omega}\sum_{j=1}^3\left[\tilde{\mathbf E}_j\right]_{w}
\sin\theta_j\partial_{x_j}
\bigg)  \nonumber 
\\
&&+\frac{e}{mc}\sum_{\alpha,\beta,\gamma=1}^3\varepsilon_{\alpha\beta\gamma}
\left[\tilde{\mathbf{B}}_\gamma\right]_{w}
\Big(\cos\theta_\beta\partial_{x_\beta}^{-1}\partial_{\theta_\beta}
-x_\beta\sin\theta_\beta-\left[\triangle\tilde{\mathbf{p}}_\beta\right]_{w}\Big)
\sin\theta_\alpha\partial_{x_\alpha}
\Bigg]w_\mathrm{g}(\mbf x,\bm\theta,t)\,,
			\label{EqModOpt}
\eea
where
\bdm
\left[\triangle\tilde{\mathbf{p}}_\alpha\right]_{w}=
-\frac{e}{mc\omega}\frac{\hbar}{i}\sum_{\beta,\gamma=1}^3
\varepsilon_{\alpha\beta\gamma}\sin\theta_\beta\partial_{x_\beta}
\int_{-1/2}^{1/2}d\tau\,\tau\mathbf{B}_\gamma
\left(\sin\theta_\sigma\partial_{x_\sigma}^{-1}\partial_{\theta_\sigma}
+x_\sigma\cos\theta_\sigma+\frac{i\hbar\tau}{m\omega}
\sin\theta_\sigma\partial_{x_\sigma},t
\right),
\edm
\bdm
\left[\tilde{\mathbf{E}}\right]_{w}=
\int_{-1/2}^{1/2}d\tau\,\mathbf{E}
\left(\sin\theta_\sigma\partial_{x_\sigma}^{-1}\partial_{\theta_\sigma}
+x_\sigma\cos\theta_\sigma+\frac{i\hbar\tau}{m\omega}
\sin\theta_\sigma\partial_{x_\sigma},t
\right),
\edm
\bdm
\left[\tilde{\mathbf{B}}\right]_{w}=
\int_{-1/2}^{1/2}d\tau\,\mathbf{B}
\left(\sin\theta_\sigma\partial_{x_\sigma}^{-1}\partial_{\theta_\sigma}
+x_\sigma\cos\theta_\sigma+\frac{i\hbar\tau}{m\omega}
\sin\theta_\sigma\partial_{x_\sigma},t
\right).
\edm
For symplectic tomogram $M_\mathrm{g}(\mbf x,\bm\mu,\bm\nu,t)$ we obtain
\bea
\partial_t M_\mathrm{g}(\mbf x,\bm\mu,\bm\nu,t)&=&\Bigg[
\frac{\bm\mu}{m}\partial_{\bm\nu}
-\frac{1}{m}\sum_{\alpha=1}^3\left[\triangle \tilde{\mathbf p}_\alpha\right]_{M}
\mu_\alpha\partial_{x_\alpha}
-e\sum_{j=1}^3\left[\tilde{\mathbf E}_j\right]_{M}
\nu_j\partial_{x_j}
\nonumber \\
&+&\frac{e}{mc}\sum_{\alpha,\beta,\gamma=1}^3\varepsilon_{\alpha\beta\gamma}
\left[\tilde{\mathbf{B}}_\gamma\right]_{M}
\Big(\partial_{x_\beta}^{-1}\partial_{\nu_\beta}
-\left[\triangle\tilde{\mathbf{p}}_\beta\right]_{M}\Big)
\nu_\alpha\partial_{x_\alpha}
\Bigg]M_\mathrm{g}(\mbf x,\bm\mu,\bm\nu,t)\,,
			\label{EqModSym}
\eea
where
\bdm
\left[\triangle\tilde{\mathbf{p}}_\alpha\right]_{M}=
-\frac{e}{c}\frac{\hbar}{i}\sum_{\beta,\gamma=1}^3
\varepsilon_{\alpha\beta\gamma}\nu_\beta\partial_{x_\beta}
\int_{-1/2}^{1/2}d\tau\,\tau\mathbf{B}_\gamma
\Big(-\partial_{x_\sigma}^{-1}\partial_{\mu_\sigma}
+i\hbar\tau\nu_\sigma\partial_{x_\sigma},t
\Big),
\edm
\bdm
\left[\tilde{\mathbf{E}}\right]_{M}=
\int_{-1/2}^{1/2}d\tau\,\mathbf{E}
\Big(-\partial_{x_\sigma}^{-1}\partial_{\mu_\sigma}
+i\hbar\tau\nu_\sigma\partial_{x_\sigma},t
\Big),
\edm
\bdm
\left[\tilde{\mathbf{B}}\right]_{M}=
\int_{-1/2}^{1/2}d\tau\,\mathbf{B}
\Big(-\partial_{x_\sigma}^{-1}\partial_{\mu_\sigma}
+i\hbar\tau\nu_\sigma\partial_{x_\sigma},t
\Big).
\edm
As it should be, equations (\ref{EqModOpt}) and (\ref{EqModSym}) 
in the classical limit $\hbar\to0$ are converted into the equations
(\ref{LivTomOpt}) and (\ref{LivTomSym}).

Combining formulas (\ref{eq_51}) and (\ref{WigNew}) we can write 
\bdm
M_{\mathrm g}(\mbf x,\bm\mu,\bm\nu)=\int\langle\mbf q|\hat\rho|\mbf q'\rangle
\langle\mbf q'|\hat U_{M_\mathrm g}(\mbf x,\bm\mu,\bm\nu)|\mbf q\rangle
d^3q\,d^3q',
\edm
where we introduced the designation for the matrix element of the corresponding
dequantizer
\bea
\langle\mbf q'|\hat U_{M_\mathrm g}(\mbf x,\bm\mu,\bm\nu)|\mbf q\rangle&=&
\frac{1}{(2\pi\hbar)^3}\prod_{\sigma=1}^3|\nu_\sigma|^{-1}
\exp\Bigg\{
\frac{i}{\hbar}(q_\sigma'-q_\sigma)\Bigg[
\frac{x_\sigma}{\nu_\sigma}-\frac{\mu_\sigma(q_\sigma'+q_\sigma)}{2\nu_\sigma}
\nonumber \\
&+&\frac{e}{c}\int_{-1/2}^{1/2}d\tau A_\sigma\left(\frac{\mbf q'+\mbf q}{2}+\tau(\mbf q'-\mbf q)\right)
\Bigg]
\Bigg\}.
\label{matrDequant1}
\eea
From (\ref{matrDequant1}) we can see that $\hat U_{M_\mathrm g}(\mbf x,\bm\mu,\bm\nu)$
is Hermitian and non-negative operator, consequently, the tomogram 
$M_{\mathrm g}(\mbf x,\bm\mu,\bm\nu)$ is real and non-negative.

From the structure of matrix element (\ref{matrDequant1}) and the fact, that the 
components of the kinetic momentum operator  
$\hat{\mbf p}=\hat{\mbf P}-\frac{e}{c}\mbf A(\hat{\mbf q})$
do not commute, it is possible to guess that the explicit expression for the dequantizer
$\hat U_{M_\mathrm g}$ looks like
\be				\label{DequantExpl1}
\hat U_{M_\mathrm g}(\mbf x,\bm\mu,\bm\nu)=
\int\frac{d^3k}{(2\pi)^3}
\exp
\left\{
i\sum_{\sigma=1}^3k_\sigma
\left[
x_\sigma-\mu_\sigma\hat q_\sigma-\nu_\sigma\hat P_\sigma
+\nu_\sigma\frac{e}{c}A_\sigma(\hat{\mbf q},t)
\right]
\right\}
\ee
Indeed, calculation of the matrix element of operator (\ref{DequantExpl1})
gives the result (\ref{matrDequant1}).

Formula (\ref{DequantExpl1})  permits to determine
the corresponding quantizer as follows:
\be				\label{QuantExpl1}
\hat D_{M_\mathrm g}(\mbf x,\bm\mu,\bm\nu)=
\left(\frac{m\omega}{2\pi}\right)^3
\exp
\left\{
i\sqrt{\frac{m\omega}{\hbar}}\sum_{\sigma=1}^3
\left[
x_\sigma-\mu_\sigma\hat q_\sigma-\nu_\sigma\hat P_\sigma
+\nu_\sigma\frac{e}{c}A_\sigma(\hat{\mbf q},t)
\right]
\right\}.
\ee
We can see that the dequantizer and the quantizer are gauge-invariant 
in the sense of transformation of type (\ref{eq6withstar}).

After calculations for the matrix element of (\ref{QuantExpl1})
we obtain
\bea
\langle\mbf q|\hat D_{M_\mathrm g}(\mbf x,\bm\mu,\bm\nu)|\mbf q'\rangle&=&
\left(\frac{m\omega}{2\pi}\right)^3
\exp\Bigg\{
\frac{ie}{c\hbar}(\mbf q-\mbf q')
\int_{-1/2}^{1/2}d\tau \mbf A\left(\frac{\mbf q'+\mbf q}{2}+\tau(\mbf q-\mbf q')\right)
\Bigg\} \nonumber \\
&&\times\delta\Big(\mbf q-\mbf q'-\bm\nu\sqrt{m\omega\hbar}\Big)
\exp\left\{i\bm\mu\left[\bm\nu\frac{m\omega}{2}-\mbf q\sqrt{\frac{m\omega}{\hbar}}\right]\right\}
\nonumber \\
&&\times\prod_{\sigma=1}^3\exp\left\{ix_\sigma\sqrt{\frac{m\omega}{\hbar}}\right\}.
\label{matrQuant1}
\eea
Using formulas (\ref{matrDequant1})  and (\ref{matrQuant1}) it is possible to check up that
\bdm
\int\langle\mbf q_2|\hat U_{M_\mathrm g}(\mbf x,\bm\mu,\bm\nu)|\mbf q_1\rangle
\langle\mbf q_1'|\hat D_{M_\mathrm g}(\mbf x,\bm\mu,\bm\nu)|\mbf q_2'\rangle
d^3x\,d^3\mu\,d^3\nu=
\delta(\mbf q_1-\mbf q_1')
\delta(\mbf q_2-\mbf q_2').
\edm

It is obvious that the corresponding dequantizer and  the quantizer for optical tomogram
$w_{\mathrm g}(\mbf x,\mbf\theta,t)$ are related with (\ref{DequantExpl1}) and
(\ref{QuantExpl1}) as follows:
\bdm
\hat U_{w_\mathrm g}(\mbf x,\bm\theta)=
\hat U_{M_\mathrm g}\left(x_\sigma,\mu_\sigma=\cos\theta_\sigma,
\nu_\sigma=\frac{\sin\theta_\sigma}{m\omega}\right),
\edm
\bdm
\hat D_{w_\mathrm g}(\mbf x,\bm\theta)=
\int\frac{|k_1|\,|k_2|\,|k_3|}{(m\omega)^3}
\hat D_{M_\mathrm g}\left(k_\sigma\sqrt{\frac{\hbar}{m\omega}} x_\sigma,
\mu_\sigma=k_\sigma\sqrt{\frac{\hbar}{m\omega}} \cos\theta_\sigma,
\nu_\sigma=k_\sigma\sqrt{\frac{\hbar}{m\omega}} \frac{\sin\theta_\sigma}{m\omega}\right)d^3k.
\edm
Due to the fact that the components of the operator $\hat{\mbf x}(\bm\theta)$ as well as
$\hat{\mbf x}(\bm\mu,\bm\nu)$ do not commute, constructed in this section
tomographic representations are not probability representations, but
they are non-negative, normalized, and gauge-independent quasi-probability
tomographic representations.

\section{\label{Sec5}Gauge-independent probability representation}
Unfortunately, introduced in the previous section gauge-independent tomographic functions
$M_{\mathrm g}(\mbf x,\bm\mu,\bm\nu,t)$ and
$w_{\mathrm g}(\mbf x,\bm\theta,t)$ are not  distribution functions of any 
physical observable. To make up for this shortcoming, we introduce
the tomographic function
$\mathfrak{M}(x,\bm\mu,\bm\nu,t)$ as the following map of the 
gauge-independent Wigner function:
\be			\label{defGIprobrep}
\mathfrak{M}(x,\bm\mu,\bm\nu,t)=\int W_\mathrm{g}(\mbf q,\mbf p,t)
\delta(x-\bm\mu\mbf{q}-\bm\nu\mbf{p})d^3q\,d^3p.
\ee
It is evident that $\mathfrak{M}(x,\bm\mu,\bm\nu,t)$ is a distribution
function of the physical observable 
$\hat x(\bm\mu,\bm\nu)=\bm\mu\hat{\mbf q}+\bm\nu\hat{\mbf p}$,
which is a scalar product of two 6-dimensional vectors $(\bm\mu,\,\bm\nu)$ and 
$(\hat{\mbf q},\,\hat{\mbf p})$. The quantity $\mathfrak{M}(x,\bm\mu,\bm\nu,t)dx$
is the probability  to have the value of the scalar operator $\hat x(\bm\mu,\bm\nu)$
within the interval between $x$ and $x+dx$ at fixed time $t$ and fixed vector $(\bm\mu,\,\bm\nu)$.

The map inverse  to (\ref{defGIprobrep}) has, obviously, the form: 
\be			\label{InvGIprobrep}
W_\mathrm{g}(\mbf q,\mbf p,t)=\left(\frac{m\omega}{4\pi^2\hbar}\right)^3
\int\mathfrak{M}(x,\bm\mu,\bm\nu,t)
\exp\left\{i\sqrt{\frac{m\omega}{\hbar}}\left(x-\bm\mu\mbf{q}-\bm\nu\mbf{p}\right)\right\}
dx\,d^3\mu\,d^3\nu.
\ee
Combining formulas (\ref{defGIprobrep}) and (\ref{WigNew}) we obtain
the expression for the matrix element of the dequantizer operator 
$\hat U_\mathfrak{M}(x,\bm\mu,\bm\nu)$ for this representation
\bea
\langle\mbf q'|\hat U_\mathfrak{M}(x,\bm\mu,\bm\nu)|\mbf q\rangle&=&
\frac{1}{2\pi\hbar|\nu_3|}\,\delta\left(\frac{\nu_1}{\nu_3}(q_3'-q_3)-(q_1'-q_1)\right)
\delta\left(\frac{\nu_2}{\nu_3}(q_3'-q_3)-(q_2'-q_2)\right)
\nonumber \\
&\times&\exp\Bigg\{
\frac{i}{\hbar}\Bigg[x\frac{q_3'-q_3}{\nu_3}
-\mu_1\frac{q_1'^2-q_1^2}{2\nu_1}-\mu_2\frac{q_2'^2-q_2^2}{2\nu_2}-\mu_3\frac{q_3'^2-q_3^2}{2\nu_3}
\Bigg]
\nonumber \\
&\times&\exp\Bigg\{
\frac{ie}{c\hbar}(\mbf q'-\mbf q)
\int_{-1/2}^{1/2}d\tau \mbf A\left(\frac{\mbf q'+\mbf q}{2}+\tau(\mbf q'-\mbf q)\right)
\Bigg\}.
\label{matrDequant2}
\eea
Taking into account expressions of dequantizer operators in previous sections 
and matrix element (\ref{matrDequant2}) we can write the explicit expression
for the gauge-invariant dequantizer $\hat U_\mathfrak{M}(x,\bm\mu,\bm\nu)$
\be			\label{Dequprobrep}
\hat U_\mathfrak{M}(x,\bm\mu,\bm\nu)=\int\frac{dk}{2\pi}
\exp\left\{ik\left[x-\bm\mu\hat{\mbf{q}}-\bm\nu\hat{\mbf{P}}
+\frac{e}{c}\bm\nu\mbf{A}(\hat{\mbf{q}})\right]\right\}.
\ee
Then the quantizer $D_\mathfrak{M}(x,\bm\mu,\bm\nu)$, obviously,  equals 
\be			\label{Quantprobrep}
\hat D_\mathfrak{M}(x,\bm\mu,\bm\nu)=
\left(\frac{m\omega}{2\pi}\right)^3
\exp\left\{i\sqrt{\frac{m\omega}{\hbar}}\left[x-\bm\mu\hat{\mbf{q}}-\bm\nu\hat{\mbf{P}}
+\frac{e}{c}\bm\nu\mbf{A}(\hat{\mbf{q}})\right]\right\},
\ee
and the calculation of it's matrix element gives rise to the following:
\bea
\langle\mbf q|\hat D_\mathfrak{M}(x,\bm\mu,\bm\nu)|\mbf q'\rangle&=&
\left(\frac{m\omega}{2\pi}\right)^3
\exp\Bigg\{
\frac{ie}{c\hbar}(\mbf q-\mbf q')
\int_{-1/2}^{1/2}d\tau \mbf A\left(\frac{\mbf q'+\mbf q}{2}+\tau(\mbf q-\mbf q')\right)
\Bigg\} \nonumber \\
&&\times\delta\Big(\mbf q-\mbf q'-\bm\nu\sqrt{m\omega\hbar}\Big)
\exp\left\{i\bm\mu\left[\bm\nu\frac{m\omega}{2}-\mbf q\sqrt{\frac{m\omega}{\hbar}}\right]
+ix\sqrt{\frac{m\omega}{\hbar}}\right\}.
\label{matrQuant2}
\eea
The correspondence rules (\ref{CorrespRules}) for representation 
$\mathfrak{M}(x,\bm\mu,\bm\nu)$ acquire a slightly modernized form
\be		\label{CorrespRules1}
\begin{array} {lcl} 
q_\sigma W_\mathrm{g}(\mathbf{q},\mathbf{p}) &\longleftrightarrow &
-\partial_{x}^{-1}\partial_{\mu_\sigma}\mathfrak{M}(x,\bm\mu,\bm\nu),
\\
p_\sigma W_\mathrm{g}(\mathbf{q},\mathbf{p}) &\longleftrightarrow &
-\partial_{x}^{-1}\partial_{\nu_\sigma}\mathfrak{M}(x,\bm\mu,\bm\nu),
\\
\partial_{q_\sigma} W_\mathrm{g}(\mathbf{q},\mathbf{p}) &\longleftrightarrow &
\mu_\sigma\partial_{x}\mathfrak{M}(x,\bm\mu,\bm\nu),
\\
\partial_{p_\sigma} W_\mathrm{g}(\mathbf{q},\mathbf{p}) &\longleftrightarrow &
\nu_\sigma\partial_{x}\mathfrak{M}(x,\bm\mu,\bm\nu).
\end{array}
\ee
With the help of (\ref{CorrespRules1}) equation (\ref{EqWigNew}) is transformed to
the evolution equation for the  tomogram $\mathfrak{M}$
\bea
\partial_t\,\mathfrak{M}(x,\bm\mu,\bm\nu,t)&=&\Bigg[
\frac{\bm\mu}{m}\partial_{\bm\nu}
-\frac{1}{m}\sum_{\alpha=1}^3\left[\triangle \tilde{\mathbf p}_\alpha\right]_{\mathfrak{M}}
\mu_\alpha\partial_{x}
-e\sum_{j=1}^3\left[\tilde{\mathbf E}_j\right]_{\mathfrak{M}}
\nu_j\partial_{x}
\nonumber \\
&+&\frac{e}{mc}\sum_{\alpha,\beta,\gamma=1}^3\varepsilon_{\alpha\beta\gamma}
\left[\tilde{\mathbf{B}}_\gamma\right]_{\mathfrak{M}}
\Big(\partial_{\nu_\beta}
-\left[\triangle\tilde{\mathbf{p}}_\beta\right]_{\mathfrak{M}}\partial_{x}\Big)
\nu_\alpha
\Bigg]\mathfrak{M}(x,\bm\mu,\bm\nu,t)\,,
			\label{EqModSymProb}
\eea
where
\bdm
\left[\triangle\tilde{\mathbf{p}}_\alpha\right]_{M}=
-\frac{e}{c}\frac{\hbar}{i}\sum_{\beta,\gamma=1}^3
\varepsilon_{\alpha\beta\gamma}\nu_\beta\partial_{x}
\int_{-1/2}^{1/2}d\tau\,\tau\mathbf{B}_\gamma
\Big(-\partial_{x}^{-1}\partial_{\mu_\sigma}
+i\hbar\tau\nu_\sigma\partial_{x},t
\Big),
\edm
\bdm
\left[\tilde{\mathbf{E}}\right]_{M}=
\int_{-1/2}^{1/2}d\tau\,\mathbf{E}
\Big(-\partial_{x}^{-1}\partial_{\mu_\sigma}
+i\hbar\tau\nu_\sigma\partial_{x},t
\Big),
\edm
\bdm
\left[\tilde{\mathbf{B}}\right]_{M}=
\int_{-1/2}^{1/2}d\tau\,\mathbf{B}
\Big(-\partial_{x}^{-1}\partial_{\mu_\sigma}
+i\hbar\tau\nu_\sigma\partial_{x},t
\Big).
\edm
In the limit case $\hbar\to0$ we get the classical equation
\begin{eqnarray}
\partial_t\,\mathfrak{M}_\mathrm{cl}(x,\bm\mu,\bm\nu,t)&=&
\left[\frac{\bm\mu}{m}\partial_{\bm\nu}
+\frac{e}{mc}\sum_{\alpha,\beta,\gamma=1}^3 \varepsilon_{\alpha\beta\gamma}
B_\gamma\Big(-\partial_{x}^{-1}\partial_{\mu_\sigma},t\Big)
\nu_\alpha \partial_{\nu_\beta} \right.
\nonumber \\
&&-\left.e\sum_{j=1}^{3}
E_j\Big(-\partial_{x}^{-1}\partial_{\mu_\sigma},t\Big)\nu_j\partial_{x}
\right]\mathfrak{M}_\mathrm{cl}(x,\bm\mu,\bm\nu,t),
		\label{LivTomSym1}
\end{eqnarray}
which is the Liouville equation in corresponding representation.
Thus, we have constructed the gauge-independent probability representation
having the clear physical meaning and the classical limit.

The density matrix $\rho(\mbf q,\mbf q',t)$ depends on time and on six spatial variables,
while the tomogram $\mathfrak{M}(x,\bm\mu,\bm\nu,t)$ depends on time, on one spatial variable, and 
on six tomography parameters.
But the number of these parameters can be reduced by one if we take into account that
the tomogram $\mathfrak{M}(x,\bm\mu,\bm\nu,t)$ is a homogeneous function
in the sense
\bdm
\mathfrak{M}(rx,r\bm\mu,r\bm\nu,t)=|r|^{-1}\mathfrak{M}(x,\bm\mu,\bm\nu,t).
\edm
Therefore, in the 6-dimensional space $(\bm\mu,\tilde{\bm\nu})=(\bm\mu,m\omega\bm\nu)$
one can, for instance, pass to the unit sphere and reduce the number of variables 
introducing the new tomogram $\mathfrak{w}(x,\bm\xi,t)$
as follows:
\be			\label{defSympgaugeInd}
\mathfrak{w}(x,\bm\xi,t)=\mathfrak{M}\left(x,\bm\mu(\bm\xi),\frac{\tilde{\bm\nu}(\bm\xi)}{m\omega},t\right)=
\int W_\mathrm{g}(\mbf q,\mbf p,t)
\delta\left(x-\bm\mu(\bm\xi)\,\mbf{q}-\tilde{\bm\nu}(\bm\xi)\,\frac{\mbf{p}}{m\omega}\right)d^3q\,d^3p,
\ee
where $\bm\xi$ is a 5-dimensional vector of directional angles in the 6-dimensional space and
\be		\label{sferCord}
\left(\begin{array} {l} 
\bm\mu(\bm\xi) \\
\tilde{\bm\nu}(\bm\xi)
\end{array}\right)
=
\left(\begin{array} {l} 
\mu_1(\xi_1,\xi_2,\xi_3,\xi_4,\xi_5) \\
\mu_2(\xi_1,\xi_2,\xi_3,\xi_4,\xi_5) \\
\mu_3(\xi_2,\xi_3,\xi_4,\xi_5) \\
\nu_1(\xi_3,\xi_4,\xi_5) \\
\nu_2(\xi_4,\xi_5) \\
\nu_3(\xi_5) 
\end{array}\right)
=
\left(\begin{array} {l} 
\sin\xi_1\sin\xi_2\sin\xi_3\sin\xi_4\sin\xi_5
\\
\cos\xi_1\sin\xi_2\sin\xi_3\sin\xi_4\sin\xi_5
\\
\cos\xi_2\sin\xi_3\sin\xi_4\sin\xi_5
\\
\cos\xi_3\sin\xi_4\sin\xi_5
\\
\cos\xi_4\sin\xi_5
\\
\cos\xi_5
\end{array}\right).
\ee
In physical meaning $\mathfrak{w}(x,\bm\xi,t)dx$
is the probability of the system to have the projection of the 
vector $(\mbf q,\,\mbf p/m\omega)$ on the direction of the
unit vector (\ref{sferCord}) within the interval between $x$ and $x+dx$.

The inverse transformation $\mathfrak{w}(x,\bm\xi,t)\to W_\mathrm{g}(\mbf q,\mbf p,t)$
has, obviously, the form:
\bea
W_\mathrm{g}(\mbf q,\mbf p,t)&=&(4\pi^2m\omega)^{-3}
\int\mathfrak{w}(x,\bm\xi,t)
\exp\left\{ir\left(x-\bm\mu(\bm\xi)\mbf{q}-\tilde{\bm\nu}(\bm\xi)\frac{\mbf{p}}{m\omega}\right)\right\}
\nonumber \\
&\times&r^5\sin\xi_2\sin^2\xi_3\sin^3\xi_4\sin^4\xi_5\,
dx\,dr\,d^5\xi.
			\label{InvGIprobrep1}
\eea
So, the tomogram $\mathfrak{w}(x,\bm\xi,t)$ also contains all available
information about the state of the system under study,
but it depends on the same number of variables as the density matrix,
and it is gauge-independent.

For the function $\mathfrak{w}(x,\bm\xi,t)$ it is also possible to write the evolution equation,
and it is possible to reduce the number of variables of $\mathfrak{M}(x,\bm\mu,\bm\nu,t)$ 
by a more symmetrical method different from (\ref{defSympgaugeInd}) - (\ref{sferCord}),
but it may be the subject of future publications.

\section{\label{Sec6}Conclusion}
In conclusion we point out that the evolution  equation of a tomogram of the state 
of quantum system, as well as the appropriate Moyal equation possess a gauge invariance.
But the optical and symplectic tomograms in their determination 
with the help of gauge-independent dequantizers (\ref{dequantizerOPT}) and (\ref{dequantizerSYMP})
do not possess the gauge independence and are converted by the integral transformation
(\ref{eq11_2}) with the kernel of type (\ref{eq11_1})
dependent on the quantizer and dequantizer operators, and the gauge function $\chi$.

Contrary to the quantum case, optical and symplectic tomograms 
of classical distribution function in the phase space with kinetic momentum possess of the gauge independence.
Therefore, in the electro-magnetic field  the evolution equations 
(\ref{EqTomOpt}) and (\ref{EqTomSym}) for gauge-independent tomograms
do not have the classical limit (\ref{LivTomOpt}) 
and (\ref{LivTomSym}) when $\hbar\to0$. 
This quality differs from the quality of the Moyal equation, which is gauge-dependent
but, nevertheless, has the gauge-independent Liouville equation as the classical limit.

To decide this problem we introduced the gauge-independent optical and symplectic 
tomographic quasi-distributions and tomographic probability distributions,
and obtained gauge-independent evolution equations for them, which in the 
classical limit are converted to the Liouville equation in corresponding tomographic 
representations.


\end{document}